\def\ltsim{\mathrel{\rlap{\lower 3pt\hbox{$\sim$}}\raise
2.0pt\hbox{$<$}}}
\def\gtsim{\mathrel{\rlap{\lower 3pt\hbox{$\sim$}} \raise
2.0pt\hbox{$>$}}}

\newcommand{\q}{\begin{equation}}
\newcommand{\qa}{\begin{eqnarray}}
\newcommand{\qs}{\begin{eqnarray*}}
\newcommand{\nq}{\end{equation}}
\newcommand{\nqa}{\end{eqnarray}}
\newcommand{\nqs}{\end{eqnarray*}}
\def\be{\begin{equation}}
\def\ee{\end{equation}}

\documentclass{aa}  

\usepackage{graphicx}

\usepackage[varg]{txfonts}

\def\lsim{\mathrel{\rlap{\lower3.5pt\hbox{\hskip0.5pt$\sim$}}
    \raise0.5pt\hbox{$<$}}}                
\def\gsim{~\rlap{$>$}{\lower 1.0ex\hbox{$\sim$}}}

\begin{document}
\title{A study on the multicolour evolution of Red Sequence galaxy populations: insights from hydrodynamical simulations and semi-analytical models.}

\author{A.D. Romeo\inst{1}, Xi Kang\inst{1}, E. Contini\inst{1}, J. Sommer-Larsen\inst{2,3}, R. Fassbender\inst{4,5}, N.R. Napolitano\inst{6}, V. Antonuccio-Delogu\inst{7} \and I. Gavignaud\inst{8}
          }

\institute{Purple Mountain Observatory, Partner Group of MPI for Astronomy, Chinese Academy of Sciences, 2 Beijing XiLu, 210008 Nanjing, China;
              \email{romeo@pmo.ac.cn}\\
\and
	Excellence Cluster Universe, Technische Universit\"at M\"unchen, Boltzmannstrasse 2, D-85748 Garching, Germany\\
\and
	Dark Cosmology Centre, Niels Bohr Institute, University of Copenhagen, Juliane Maries Vej 30, DK-2100 Copenhagen, Denmark\\
\and
	INAF - Osservatorio Astronomico di Roma, Via Frascati 33, I-00040 Monteporzio Catone, Italy\\
\and
        Max-Planck-Institut f\"ur extraterrestrische Physik (MPE), Postfach 1312, Giessenbachstr.,  85741 Garching, Germany\\
\and
	INAF - Osservatorio Astronomico di Capodimonte, salita Moiariello 16, I-80131 Napoli, Italy\\
\and
        INAF - Osservatorio Astrofisico di Catania, via S.Sofia 78, I-95123 Catania, Italy\\
\and
	Departamento de Ciencias Fisicas, Universidad Andres Bello, Av. Republica 220, Santiago, Chile
             }

   \date{Received; accepted}

\abstract
{By means of our own cosmological-hydrodynamical simulation (SIM) and semi-analytical model (SAM)
we studied galaxy population properties in clusters and groups, spanning over 10 different bands from UV to NIR, 
and their evolution since redshift $z$=2.}
{We compare our results in terms of galaxy red/blue fractions and luminous-to-faint ratio (LFR) on the Red Sequence (RS)
with recent observational data reaching beyond $z$=1.5.}
{Different selection criteria were tested in order to retrieve the galaxies effectively belonging to the RS: either by their quiescence degree 
measured from their specific SFR (``Dead Sequence''), or by their position in a colour-colour plane which is also a function of sSFR. 
In both cases, the colour cut and the lower limit magnitude thresholds were let evolving with redshift, in order to follow the natural shift of the characteristic luminosity in the LF. }
{We find that the Butcher-Oemler effect is wavelength-dependent, with the fraction of blue galaxies increasing steeper in optical-optical 
than in NIR-optical colours. Besides, a steep trend in blue fraction can be reproduced only when an optical fixed luminosity-selected sample is chosen,
while it flattens when selecting samples by stellar mass or an evolving magnitude limit.
We then find that also the RS-LFR behaviour, highly debated in the literature, is strongly dependent on the galaxy selection function: 
in particular its very mild evolution recovered when using a mass-selected galaxy sample, is in agreement with values reported for some of the highest 
redshift confirmed (proto)clusters.
As to differences through environments, we find that normal groups and (to a lesser extent) cluster outskirts
present the highest values of both star forming fraction and LFR at low $z$, while fossil groups and cluster cores the lowest:
this separation among groups begins after $z\sim0.5$, while at earlier epochs all group star forming properties are undistinguishable.}
{Our results support a picture where star formation is still active in SIM galaxies at redshift 2, in contrast with SAM galaxies that
have formed earlier and are already quiescent in cluster cores at that epoch. Over the whole interval considered, we also find that the more massive 
RS galaxies from the mass-selected sample grow their stellar mass at a higher rate than less massive ones. On the other side no dearth of red dwarfs is reported at 
$z\gsim$1 from both models. }

\titlerunning{Multicolour evolution of RS galaxies}
\authorrunning{A.D. Romeo et al.}
\maketitle

\section{Introduction}

The red/blue fraction and the giant-to-dwarf or luminous-to-faint ratio are common proxies that contribute to characterize two fundamental
properties of the galaxy populations in clusters or groups: the Red Sequence (RS) of early-type (ET) galaxies in the colour-magnitude plane, 
and the luminosity function (LF), respectively.
As to the latter, its characteristic luminosity $L^*$ is usually taken as a measure of the mean luminosity of giant
galaxies, while its slope measures the relative abundance of dwarf galaxies. The variation of these two parameters with
redshift and environment allows to estimate the growth of the two populations, either in terms of stellar mass or light.
In particular the LF of RS galaxies (``RSLF'') has been extensively studied, based on which it has been reported a deficit of faint red galaxies at $z\sim 1$ and beyond, 
both in optical and in redder colours (De Lucia et al. 2007, Krick et al. 2008, Gilbank et al. 2008, Stott et al. 2009, Hilton et al. 2009): this would imply 
that less massive galaxies undergo a slower evolution inasmuch as their star forming activity lasts longer with respect to more massive ones.
Nevertheless, works by Andreon (e.g. 2008, 2014) and Crawford, Bershady \& Hessel (2009) point towards the opposite direction, highlighting a non-evolution
of the faint end slope of the LF. 
The dependence of the latter on environmental and evolutionary parameters is still debated, as is the presence of an upturn at faint magnitudes 
(see Zucca et al. 2009, Ba\~nados et al. 2010, de Filippis et al. 2011).
The controversy is complicated by the fact that different filters, apertures and limiting magnitudes are used in the literature (see Section 3).
To this respect stellar mass-based surveys better allow to study the evolution of galaxy populations independently of colours and magnitudes,
providing a more straightforward parallel to theoretical models: recently a study on field galaxies by Tomczak et al. (2014) measured a steepening of the 
galaxy Stellar Mass Function's slope at its low-mass end, with an upturn at masses $<10^{10}M_{\odot}$ extending to $z=$2 and beyond.

On the other side, the red or the blue fractions are directly connected to the slope and scatter of the RS, that in turn provide relevant constraints upon the epoch and duration of the star formation in ETGs. In fact, the RS itself is shaped also according to changes in the specific SF rate (SFR) of member galaxies, as they move from a previously populated ``blue cloud''. 
In Romeo et al. (2008) we had shown that the epoch when the switch occurs between the star-forming and the passive regime is closely mirroring
the epoch at which the RS slope gets null, that is also when its scatter becomes that of the pure quiescent population.
Such epoch of migration towards the RS and subsequent passive reddening coincides then with shutting off the bulk of SF: this may be in turn determined by environmental input like (gas-poor) mergers between galaxies, or by a natural mass threshold set up by internal sources hindering the cold phase condensation, like energy feedback from AGN (see Cattaneo et al. 2008).
The role of AGN feedback helps preventing the cooling of hot gas around massive galaxies, and the subsequent formation of a new disc.
Nevertheless Kang, van den Bosch \& Pasquali (2007) have suggested that high and low-mass ETGs formed out of dry and wet mergers, respectively,
independent of the presence of an AGN.
A comprehensive analysis carried on by Peng et al. (2010), and later extended by Raichoor \& Andreon (2012b), demonstrated that the red/blue fraction is a
composite yet separable function of both redshift, environment (as measured by the cluster-centric distance in clusters, or by the galaxy density in general)
and galaxy stellar mass as well, leading to two quenching modes that act as independently: one driven by environment and another by galaxy mass, only the
latter of which is dependent on redshift.

The issue of star formation at high redshift is widely studied in clusters, that nonetheless suffer from some selection bias: this occurs for example
when they are selected by means of the Red Sequence itself, that then becomes a premise {\it a priori} rather than a diagnostic tool; the opposite may happen
in the field, where blue bright galaxies are more likely to be detected at high redshift.
Many surveys have measured a significant growth in mass of the RS since $z\sim$1 (e.g. Gladders et al. 1998, Bell et al. 2004, Mei et al. 2009),
generally basing their results on {\it U} to {\it I}-centered filters.
On the other hand, Bundy, Ellis \& Conselice (2005), among others, have found little evolution in the distribution of
massive galaxies since $z\sim$1 to present, using {\it K}-band to constrain the stellar mass.
Lidman et al. (2008) studied one cluster at $z$=1.4 in {\it J-K}, finding an already well defined RS in the core,
with a scatter in colours around 0.055, whereas galaxies in the outskirts appear younger and bluer, half of them with high SFR ongoing.
More recently Fassbender et al (2014) measured a bright end of the LF still evolving at $z\simeq$1.6, with a very active ongoing mass-assembly and
clear merger signatures for the most massive galaxies, along with a RS well populated at magnitudes fainter than $K_s^*$ but still lacking bright objects 
above it.
A significant population of massive galaxies much bluer than the RS is detected also by Strazzullo et al (2013), who probed the massive end galaxies in 
a proto-cluster's core at $z$=2, finding therein a mixture of both quiescent and star-forming. Contrasting results from Raichoor \& Andreon (2012a),
Andreon et al. (2014) and Newman et al. (2014) indicate instead that all most massive core galaxies are already red and passive in a proto-cluster
spectroscopically confirmed between $z$=2.2 and 1.8, having a deep completeness in galaxy mass.

As a widespread manifestation of the scenario above, the Butcher-Oemler (BO) effect (Butcher \& Oemler 1984 = BO84) implies that clusters at 
higher redshifts have higher fraction $f_{bl}=N_{blue}/N_{tot}$ of blue galaxies.
Since that pioneer photometric work, that was valid up to $z$=0.6, galaxy infall has turned out as a likely mechanism for increasing the blue fraction in 
intermediate redshift clusters (e.g. Kauffmann 1995; Couch et al. 1998; van Dokkum et al. 1998; Ellingson et al. 2001; Fairley et al. 2002).
Indeed a time declining blue fraction in the hierarchical framework is a natural consequence of environmental suppression of star formation
in overdense systems (see Diaferio et al. 2001), whereby the transition from blue to red regimes occurs first and mainly outside the cluster core
in response to processes such as ram pressure stripping, galaxy encounters, or strangulation (e.g. Berrier et al. 2009).
Evidences that the RS scatter was still tight at higher redshift (e.g. Ellis et al. 1997, at $z$=0.55; Stanford,
Eisenhardt \& Dickinson 1998, at $z$=0.9), imply a homogeneity of the ET population across the cosmic time which is
actually in contrast with the strong evolution of the cluster galaxies predicted by some interpretations of the BO effect.
The latter can be interpreted by means of the transformation of blue field galaxies into red cluster ones, but only if such infall is
accompanied by a morphological change from spirals to ETs (or an equivalent change from star-forming to passive), 
then the scatter of the CMR is expected to increase, as a result of the higher number of blue ETs approaching the RS
(see Schawinski et al. 2014, Cen 2014). Moreover, a recent ($z<$1.3) transformation of many blue galaxies in faint red ones would modify the 
faint-end slope of the LF itself and again increase the RS scatter to an extent that is not observed at that epoch (see Andreon 2008).

Thus the evidence of the BO effect is under many aspects still controversial, due to discrepancies
in the sample selections (waveband, completeness limit, cluster dynamics) and defining criteria for a red/blue colour threshold.
Selection of galaxy samples in the near-infrared as opposed to the optical should result in samples more representative by 
stellar mass (Arag\'on-Salamanca et al. 1993; Gavazzi, Pierini \& Boselli 1996), as NIR or MIR colours are almost insensitive
to star formation  histories.
Given that luminosities and colours of galaxies selected in the {\it K} band more reliably trace their behaviour in terms
of stellar mass, {\it K}-selected samples can be used to constrain the masses of the blue galaxies 
in excess at high $z$, which are thought to evolve into S0 galaxies or dwarfs later on (De Propris et al. 2003).
Indeed Haines et al. (2009) found that the star-forming fraction in {\it J-K} follows a fairly mild evolution at $0<z<0.4$,
that becomes constant at around 50\% if considering only the innermost galaxies within $R_{500}$; this indicates that
the residual BO effect would be entirely due to the infall galaxies in the outer cluster regions. 

In this paper we aim at modelling, by means of cosmological and hydrodynamical simulations (hereafter SIM) including chemical evolution,
and with the independent aid of a novel semi-analytical model (SAM), which are both described in Section 2,
the global properties of the ET galaxies' assembly and evolution in over-dense regions. 
This will be pursued by reproducing the evolution of the RS luminous-to-faint ratio (LFR) and the blue fraction,
as complementary yet independent means of diagnostics.
In order to shed light upon the combined mechanisms that are at the base of the regularity in the global properties 
of the ET populations in clusters and groups, one important step would be to choose reasonable criteria for defining
a correspondence between the observed RS and a subsample of ET galaxies descending from the models that, by their ages and metallicities,
can be safely considered as belonging to the RS: this task will be detailed in Section 3.
The motivation behind the choice of our diverse model sets is then to assess how the different methodologies with respect
to observational data affect our results, as well as to investigate their dependence on the physical processes implemented in either model.

\section{Methods}

\subsection{The simulations}
The formation of our galaxies is followed ab initio within a 150 Mpc cosmological volume, that allows to account for their interaction with a large scale environment. The initial N-body simulation adopts a standard $\Lambda CDM$ cosmology, with $\Omega_\Lambda$=0.7, $\Omega_m$=0.3$, \Omega_{b}$=0.045,
$h$=0.7, and $\sigma_8=0.9$.
This simulation serves as input for the volume renormalization technique to achieve higher resolution in zoomed
regions such as individual clusters and groups, that are later re-simulated with the full hydrodynamical code (see Romeo et al. 2006). 
Thus the mutual cycle between IGM, star formation (SF) and stellar feedback  can be described in a self-consistent way at the cluster scale, by means of the baryonic physics implemented in the hydrodynamical code: this includes a metal dependent radiative cooling function (with cooling shut-off below $10^4$K),
thermal conductivity, star formation according to a top-heavy IMF, chemical evolution with not-instantaneous recycling, feedback from SN-Ia and SN-II, and SN-II driven galactic winds. The whole cycle gas-stars follows the birth and evolution of the star particles and their final “decay” again into gas particles through feedback mechanism; the latter will in turn regulate further episodes of SF, and so forth.

By re-zooming on target-selected regions at higher resolution, we were able to resolve down to galaxy-sized haloes and model their stellar populations:
a mass resolution of stellar and gas particles down to to $3\times 10^7 h^{-1}M_\odot$ was achieved by this way.
In particular the virial mass of the haloes selected at $z=0$ to be re-simulated at higher resolution are $1\times 10^{14} M_\odot$ for groups,
$3\times 10^{14} M_\odot$ for the lower mass cluster and $1.2\times 10^{15} M_\odot$ for the larger one -measured within the virial radius defined as enclosing an overdensity of 200 times the critical background density. Out  of the 12 re-simulated groups at z=0, 4 turned out to be of fossil nature (see D'Onghia et al. 2005).
Galaxies from homogeneous density regions have been stacked together, to make up four environmental classes: cluster cores (IN), cluster outskirts (OUT),
normal groups (NG) and fossil groups (FG); the separation between these cluster regions is given by one third of $R_{200}$.

In the SPH scheme, each star particle represents a Single Stellar Population (SSP) of total mass corresponding to the stellar mass resolution of the simulation.
The individual galaxies are assigned a luminosity as the sum of the luminosities 
of its constituent star particles, in the broad bands {\it UBVRIJHK} (Johnson/Cousins filter, Vega system).
The individual stellar masses are distributed according to an Arimoto-Yoshi IMF; 
each of these SSPs is characterized by its age and metallicity, from which luminosities are computed by mass-weighted integration of the Padova 
isochrones (Girardi et al., 2002). Thus the physically meaningful quantities of our data are age and metallicity, from which colours are derived.
Results from this approach have already involved the study of global galaxy properties such as the RS (Romeo et al. 2005, 2008)
and the mass-metallicity-SFR relations (Romeo Velon\`a et al. 2013).

\subsection{The S.A.M.}

The SAM used in this paper is based on Kang et al. (2005), later further developed in Kang et al. (2012), which we refer the reader to for details. 
The main ingredients are hereby introduced as following. The simulation was performed using the Gadget-2 code 
(Springel 2005) with cosmological parameters adopted from the WMAP7 data release (Komatsu et al. 2011), namely: $\Omega_{\Lambda}$=0.73, $\Omega_{m}$=0.27, 
$\Omega_{b}$=0.044, $h$=0.7 and $\sigma_{8}$=0.81; the cosmological box had a size of 200$Mpc/h$ on each side, and was populated with $1024^{3}$ particles.
The merger trees are constructed by following the subhaloes resolved (by using {\it SubFind}: Springel et al. 2001) in FOF haloes at each snapshot. The SAM is then grafted on the merger trees and self-consistently models the physics processes governing galaxy formation, such as gas cooling, star formation, supernova and AGN feedback. Finally, the galaxy luminosity and colours are calculated based on the stellar population synthesis of Bruzual \& Charlot (2003) adopting a Chabrier stellar IMF (Chabrier 2003). 

In particular the SAM includes ``radio-mode'' AGN feedback, parametrized in a phenomenological way as in Croton et al. (2006). This mechanism works as to suppress the
cooling in massive haloes, whose star formation in the central galaxies is hence shut off, drifting them onto the RS.
In the present model, developed in Kang, Jing \& Silk (2006), the heating efficiency of the surrounding gas is proportional to a power of the virial velocity of the host halo, hence the gas cooling keeps on forming stars until a massive spheroid forms at the galaxy centre. In this way more massive and luminous galaxies
can be formed at high redshift, thus achieving a successful reproduction of the observed LF in rest-frame {\it K-}band and the galaxy colour distributions.

Moreover, it is worth briefly discussing here the treatment of gas stripping from dwarfs/satellites in the model. To this regard, a common outcome of most SAMs has been the overabundance of red faint galaxies produced, that has generally been ascribed to the low efficiency in tidal disruption of dwarfs {\it tout court} (see Weinmann et al. 2011) and at the same time to a too efficient ram-pressure stripping of those survived. In particular, most SAMs assume that the entire hot gas reservoir of a galaxy is instantaneously stripped at the very event of accretion into a larger halo, that is inmediately after infall when it becomes a satellite. In our model instead we let the outer hot gas in dwarfs to be gradually stripped over a timescale of about 3 Gyr: such prolongued ``strangulation'' results in a longer lasting cooling and hence delayed quenching of SF. As demonstrared in Kang \& van den Bosch (2008) this recipe can effectively lower the dwarf fraction on the RS and, if combined with a stronger tidal disruption prior to the central accretion, can fairly well reproduce the observed RSLF without affecting its bright end.
Finally, in our SAM there is no ram-pressure stripping of cold gas from the host halo (a mechanism leading to the so-called ``starvation''), so this inner
component is normally consumed by star formation and SN feedback. 

From the initial simulation box, a total number of haloes above the threshold of $1\times 10^{14}M_{\odot}$ were extracted, ranging from 2830 ($z$=0) to 2150 ($z$=0.5), 1300 ($z$=1), down to 300 ($z$=2); out of them those below $1.5\times 10^{14}M_{\odot}$ are classified as groups.
Finally and relying on Menci et al. (2008), we note that the different normalization of the power spectrum assumed in our models, although affecting the cluster abundance at a given redshift, however is not effective in changing the dynamical history of DM haloes, nor hence their stellar mass assembly.
The colour output of galaxies and the luminosity function are instead more prone to be affected by different choices of the IMF (see Romeo, Portinari \& Sommer-Larsen, 2008), even though both our options are top-heavy.

\begin{figure}
\includegraphics[width=9cm,height=11cm]{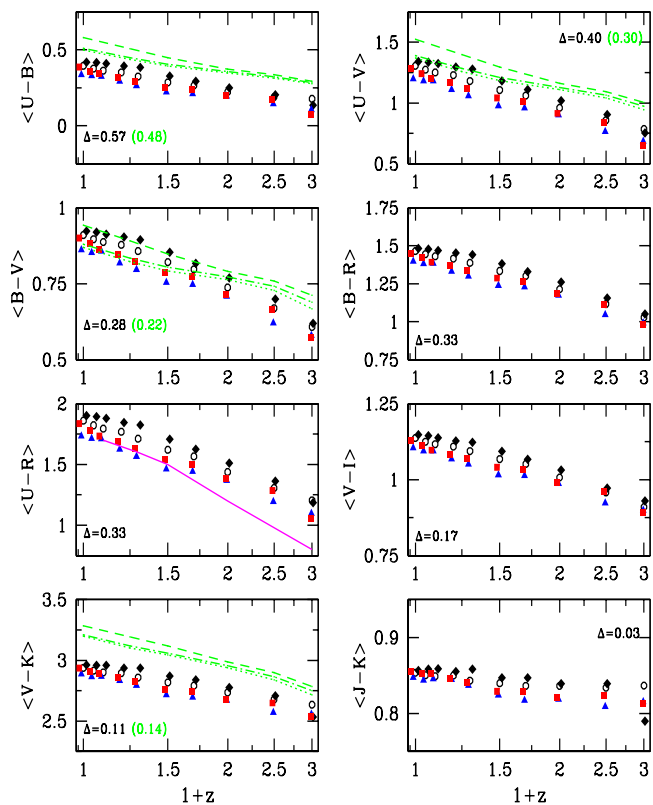}
\caption{Redshift evolution of the mean rest-frame colours of the DS samples from the SIM: clusters IN ({\it black rhombi}), clusters OUT ({\it open circles}), 
normal groups ({\it blue triangles}), fossil groups ({\it red squares}); and from the SAM ({\it green}): clusters IN ({\it dashed lines}), clusters OUT
({\it dotted lines}), and groups ({\it dot-dashed lines}). It is also indicated the difference in colour during the redshift interval: $\Delta\equiv \frac{\Delta col}{col(z=0)}$, averaged over the environments. Points are slightly offset in x-axis for clarity. This mean colour blueshifted by 0.2 dex gives our first criterion for selecting red galaxies. The synthesis model with $Z=Z_\odot$, $z_f$=5, $\tau=3.7 Gyr$ is shown in the {\it U-R} panel as magenta line (see text).}
\label{col}
\end{figure}

\begin{figure*}
\includegraphics[width=15cm,height=15cm]{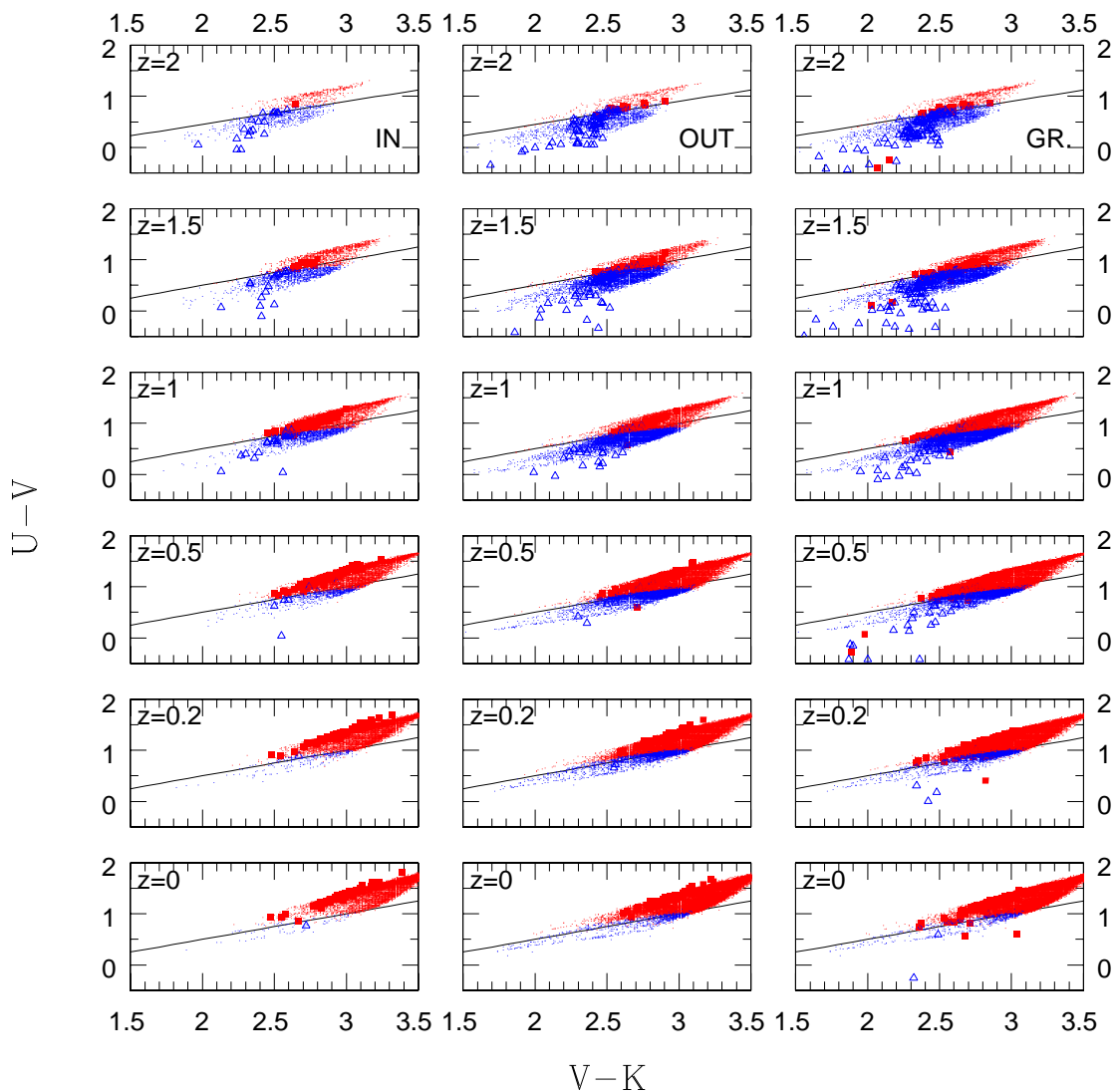}
\caption{Double-colour diagrams for galaxies more massive than $3\times10^9M_{\odot}$, selected by their specific SFR: higher ({\it blue}) and lower ({\it red})
than $10^{-11}/yr$; larger symbols are for SIM, dots for the SAM. Solid lines are given by the fitting formulae that separate the two regions:
$U-V=0.5(V-K)-0.5$ for $0\leq z<1$, $U-V=0.5(V-K)-0.6$ for $1\leq z\leq 1.5$, $U-V=0.45(V-K)-0.54$ for $1.5<z\leq 2$.}
\label{colcol}
\end{figure*}

\section{Definition of RS samples}

All models have to face up to substantial difficulties when directly confronted to observations: first, galaxy SEDs are often not well sampled with the current bands, especially at high $z$, therefore any attempt to convert to rest-frame colours would introduce in these cases a large systematic error.
Secondly, when considering the stellar masses, the dominant systematic error comes from the choice of the IMF, which gives a factor two uncertainty in the LFR results and is likely the main source of scatter in the compilation of different data from literature; our AY IMF produces stellar masses between a
Chabrier and a Salpeter IMF, that are commonly used for the conversion.
Third, all background subtracted quantities are measured within a cluster-centric radius that is quite small in the high-$z$ cases, in order to provide the highest signal-to-noise (for example the 30'' aperture corresponds to 255kpc in the $z=1.58$ cluster of Fassbender et al. 2014) -and therefore can be
consistently compared only with our cluster core points. Finally clusters themselves compose an utterly heterogeneous family and many features of galaxy populations may partially depend on correlations between cluster and galaxy properties, such as cluster's mass and dynamical state: high-$z$ surveys are
generally biased towards richer clusters, hence more dynamically evolved, which in turn will affect galaxy colours and star formation histories; evidence
of it is that lower galaxy blue fractions are measured when including X-ray selected clusters, which are on average more massive and relaxed (see Andreon \& Ettori 1999).

Throughout the following, all analysis will refer to galaxy samples excluding the BCG. All model apertures are differential, being fixed to the virial radius at each epoch,
and therefore are naturally accounting for the cluster's cosmic evolution leading to some extent of blue segregation outside the cluster cores (see Ellingson et al. 2001).

We define the ``Dead Sequence" (DS) sample as that of galaxies with no star formation ongoing over the last Gyr, and plot in Fig. \ref{col} its average colours
as a function of $z$ in different wavebands, motivated by the LF of RS galaxies being strongly dependent on the latter, as stated e.g. by Goto et al. (2005).
As a general trend, differences among the four environments are
scarce in absolute value, being more evident in {\it U-V}, {\it B-V} and {\it U-R}: the reddest colour in any band and at every $z$ 
is presented by cluster core galaxies, while normal groups have on average the bluest colours. 
This is in line with the environmental sequence found in R08, where normal groups were characterized by intrinsically higher SFR 
with respect to FGs; and also confirms that, on the opposite end, galaxies in denser cluster cores are redder.
Differences between SIM and SAM tell that the latter yields a redder colour distribution at any epoch, and especially in {\it V-K}: this is expected, 
since most SAMs are producing too many galaxies whose stellar mass bulk was formed too early (see Discussion). 
As to redshift evolution, it appears to strongly depend on the wavelength: the steepest reddening in time is
observed in the five colours within {\it UBVR} bands, where it amounts on average to one third with respect to the present colour; and especially in {\it U-B}, 
where the difference peaks at around half of this value, in stark contrast with suggestions by Willmer et al. (2006) for clusters at $z\sim1$. 
On the opposite, it is much flatter in {\it V-I} and {\it V-K} and gets almost exactly constant
(at a value below 1) in {\it J-K}. The latters are also those where the least difference is noted among environments.
This confirms that measures of optical-optical colours are strongly affected by recent episodes of star formation, whereas
on the contrary optical-NIR colours are mostly sensitive to intrinsic changes in metallicity and stellar mass.
We will use in the following these mean colours (shifted 0.2 mag bluewards) to select the RS samples as our first criterion.  
This reproduces the original definition chosen by BO84, where cluster galaxies bluer than 0.2 mag in {\it B-V} 
with respect to the RS ridge were selected.
It is worth stressing that the threshold yielded by this criterion is naturally evolving with $z$: this procedure is equivalent to
that of Andreon et al. (2006), who adopted instead an evolving colour cut $\Delta(B-V)$ below a fixed RS upper limit.
It is also consistent with the method of Raichoor \& Andreon (2012b), who assume a synthetic model from Bruzual \& Charlot (2003) computed for solar metallicity,  formation redshift of $z_f$=5 and an exponentially declining star formation history $\propto exp(-t/\tau)$ with $\tau=3.7 Gyr$, as a divisory 
line for red/blue classification at any redshift (shown in Fig. \ref{col} as reference in the {\it U-R} panel).

In Fig. \ref {colcol} our galaxies are plotted within a particular colour-colour plane that has been proven to
efficiently discriminate between quiescent and star-forming galaxies up to $z\sim2$ (see Williams et al. 2009). 
As seen when the two populations are classified by their specific SFR (sSFR), such a bi-colour plot allows to separate
them by means of simple diagonals that are almost unchanged over the whole redshift interval, and that can provide us with
an alternative selection criterion for defining the two samples: from our joint SIM+SAM distribution, 
$U-V=\alpha(V-K)-\beta$ is best fitted by
the values $(\alpha, \beta)=(0.5,0.5), (0.5,0.6), (0.45,0.54)$ respectively for $z=0-1$, $z=1-1.5$ and $z=2$.

We also checked for alternative criteria to define the RS samples and computed the {\it U-V} and {\it U-B} blue fractions following the empirical thresholds given in Bell et al. (2004) and Willmer et al. (2006), respectively -both holding for $z\leq$1:
\begin{equation}
U-V=1.15-0.31 z -0.08(M_V -5log h +20)
\label{eqUV}
\end{equation}
\begin{equation}
U-B=0.454-0.032(M_B+21.52)-0.25-0.1(z-1)
\label{eqUB}
\end{equation}
These colour cuts have been modified as luminosity-dependent, to avoid excluding faint RS galaxies due to the very slope of the RS (see Crawford et al. 2009).
Moreover, in estimating both the blue/red fraction and the LFR, an evolving limit on absolute magnitude should be applied in order to track the same
galaxy population at different redshifts (see Andreon, Lobo \& Iovino, 2004). In fact, especially in surveys covering broad redshift ranges, imposing a fixed cut
in apparent magnitude would introduce an unreasonable selection bias against redder objects at higher redshifts; nonetheless, applying a fixed threshold in visual absolute magnitude would not help enough either, because of redshift-dependent incompleteness limits. Therefore ensuring the selection of a uniformly sampled region in the colour--magnitude space can only be accomplished by following the linear redshift evolution of the characteristic luminosity of the LF at a 
given $z$.
Specifically, in {\it B}-band we adopted the threshold
proposed by Gerke et al. (2007) for groups at $z\simeq1$: 
\begin{equation}
M_B^{lim}=5log h-20.7-1.37(z-1)$=$-20.1-1.37z, 
\end{equation}
whereas in {\it V} we let evolving for analogy the initial BO $M_V$=-19.3 (corresponding to their original -20 in our cosmology) as 
\begin{equation}
M_V^{lim}=5log h-19.9-1.37(z-1)=-19.3-1.37z. 
\end{equation}
As for the {\it K}-band, we have tried both the fixed threshold $M_K^{lim}=$-22, consistent with the constancy of $M^*_K$ over redshift as measured by Kashikawa et al. (2003), and the time-dependent expression 
\begin{equation}
M_K^{lim}=-22.16-(z/1.78)^{0.47}
\end{equation}
following instead Cirasuolo et al. (2010).
All these values correspond roughly to $M^*$+2 in their respective filters, approximating the original limit of 1.8 mag below $M^*$ adopted by BO84.
However, it has to be noted that this is not the same one used for estimating the LFR, whose lower limit is taken as $M^*$+3 (see below): this means that a 
considerable population of faint galaxies comprised in this one-magnitude bin is not considered when computing the blue/red fractions,
yet still contributing to make up the LFR.
These results will be compared in the next section with those obtained by applying a lower limit in terms of stellar mass, which better helps to even up differences due to diverse passbands and completeness limits used, although at expenses of larger uncertainty from the light-to-mass conversion.

\section{Results}

\begin{figure}
\includegraphics[width=8cm,height=6cm]{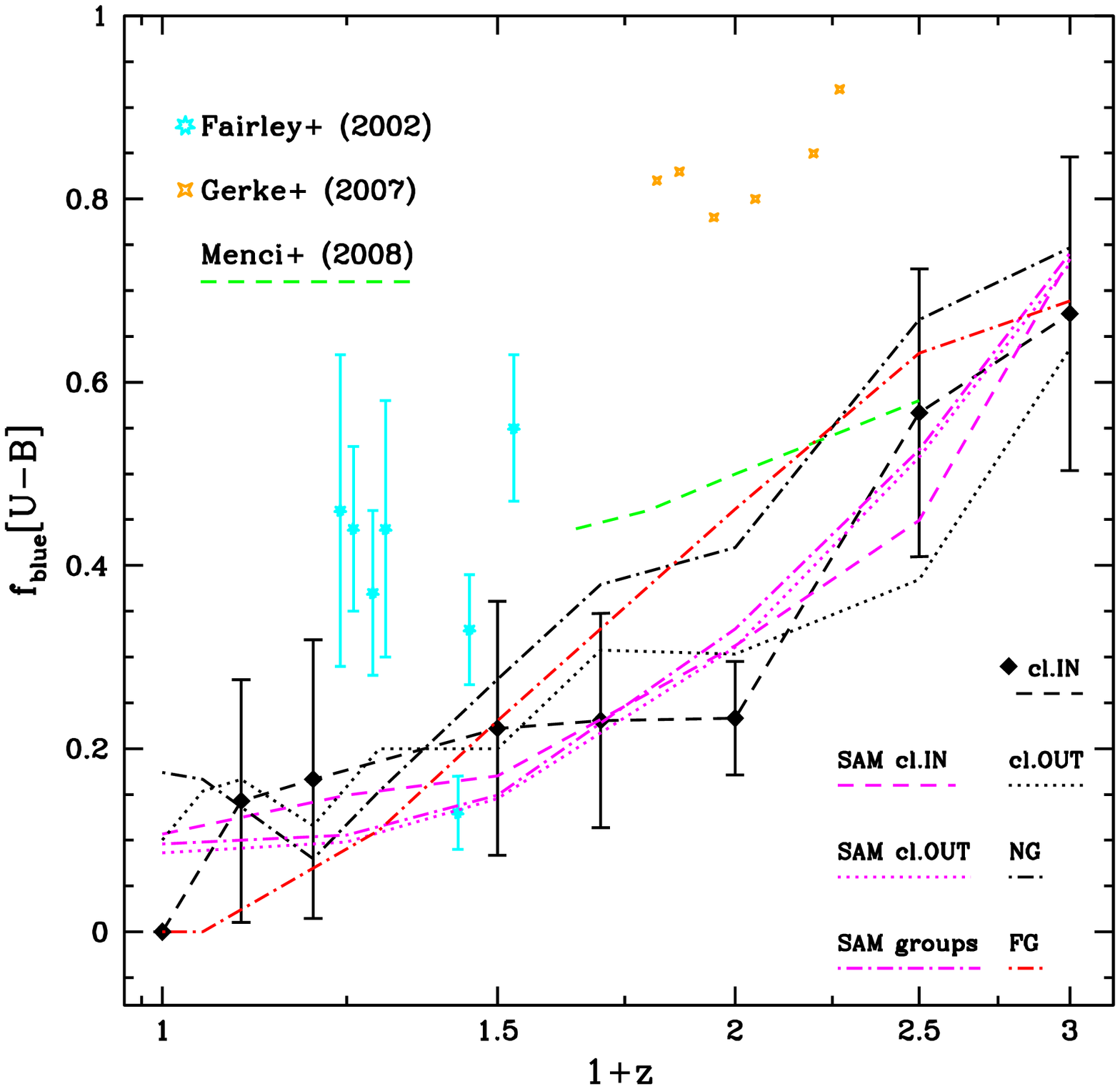}
\vfill
\includegraphics[width=8cm,height=6cm]{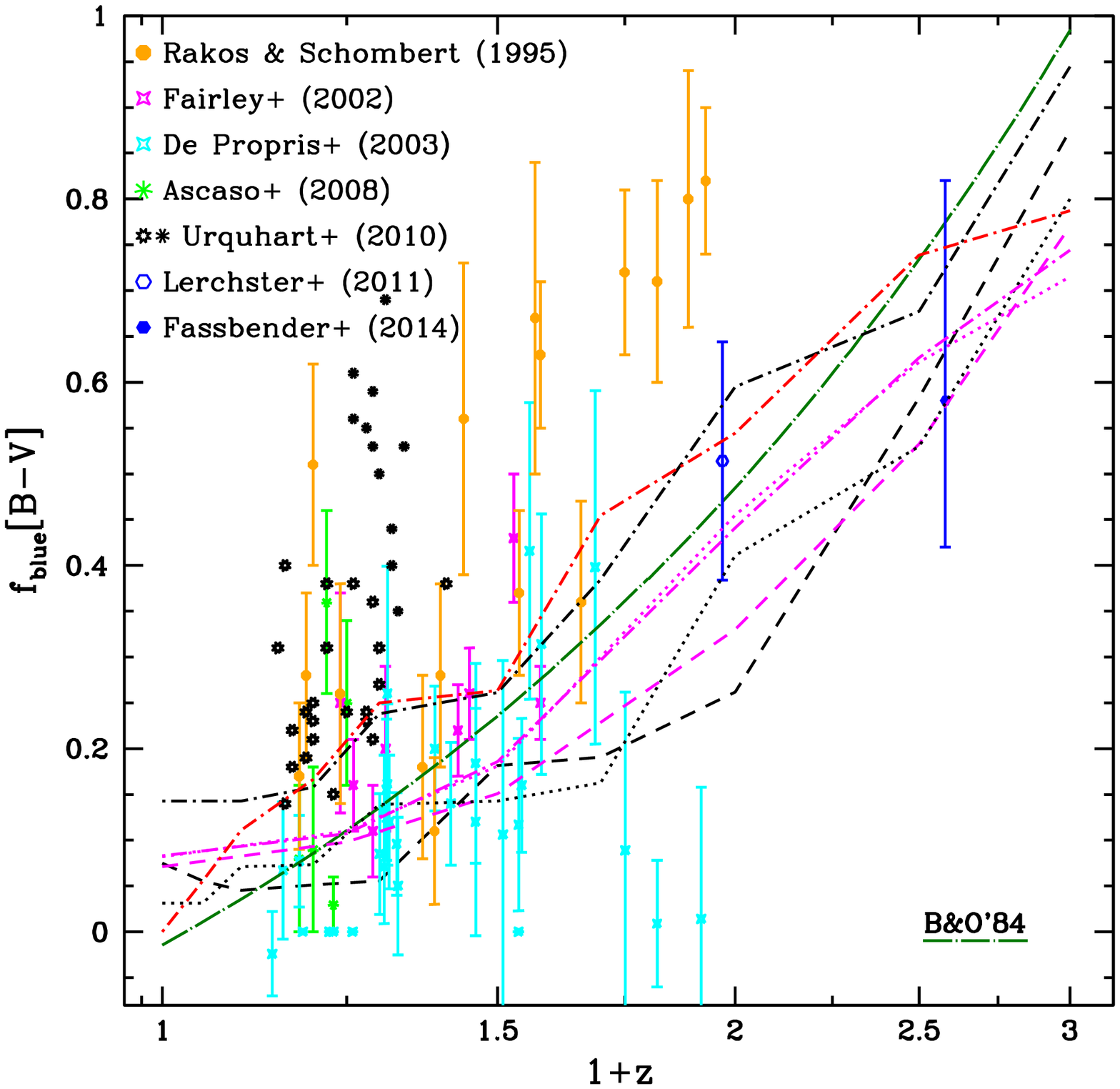}
\vfill
\includegraphics[width=8cm,height=6cm]{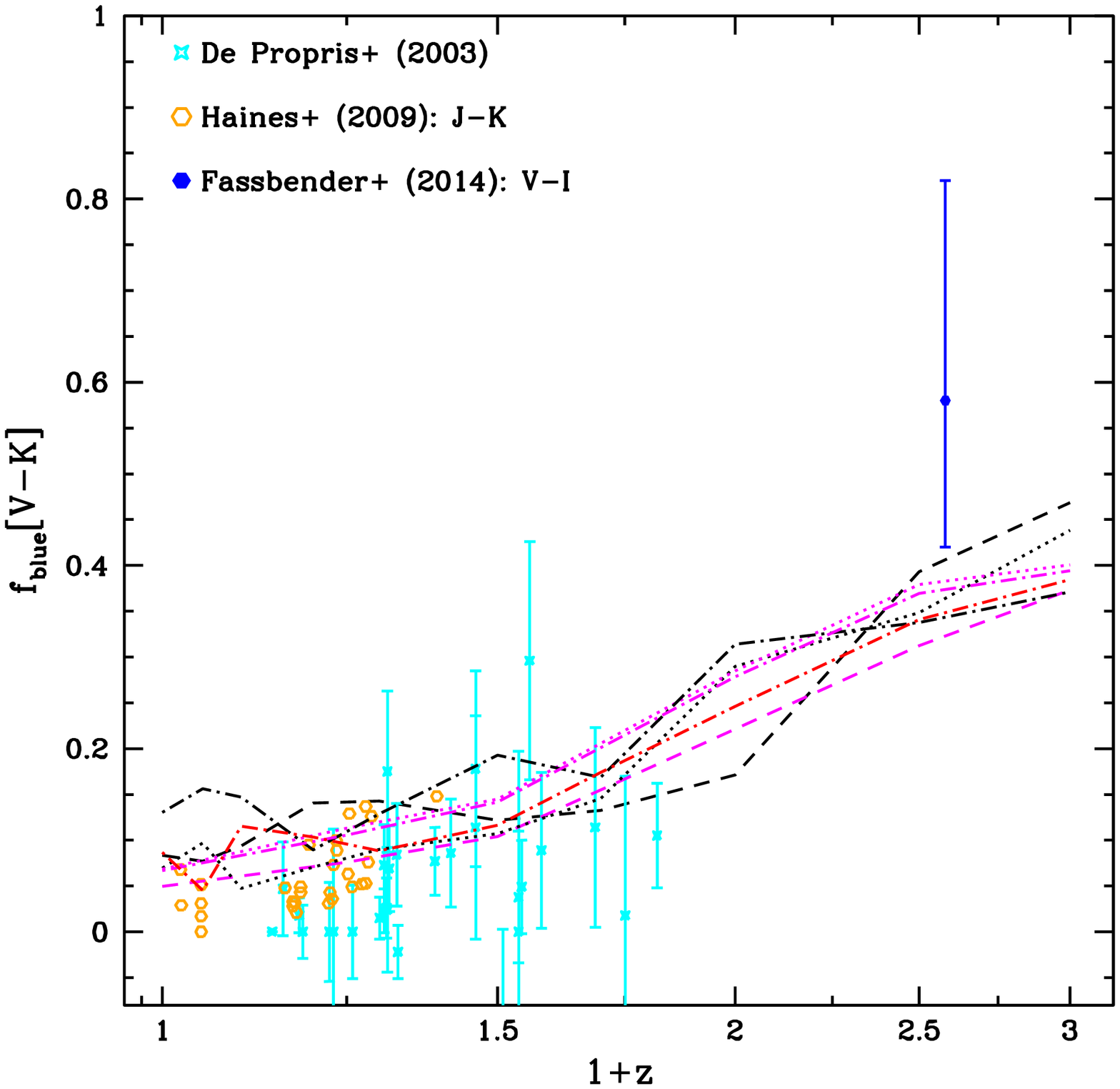}
\vfill
\caption{Fraction of blue galaxies as computed in different colours, based on the DS definition from Fig.\ref{col} and colour-colour selection from Fig.\ref{colcol}.
It is compared with observational data and with the original Butcher-Oemler in {\it B-V}, extrapolated to high $z$ as
$f_{bl}=(z/2)-0.015$ (mid panel, {\it dot-long-dashed line}). The lower limit for selecting samples is given in terms of 
redshift-variable absolute $M_B, M_V$ and $M_K$ magnitudes (see text for details).
The Poissonian error on $f_{bl}$ is given by $\sigma^2(f_{bl})=N_{bl}(N-N_{bl})/N^3$ (see De Propris et al. 2004) and shown only for SIM cluster cores 
in the {\it U-B} plot as reference (upper panel). 
}
\label{fblu}
\end{figure}

\subsection{The red and blue fractions}

Two main methodologies are commonly applied when measuring the BO effect: either relying on optically selected samples down to a rest-frame
$M_V$ equivalent to the original BO84 work (e.g. Rakos \& Schombert 1995, Fairley et al. 2002, Ascaso et al. 2008, Urquhart et al. 2010, Lerchster et al. 2011), 
or selecting in observed $K$-band applying a brightness limit fixed with respect to the characteristic $K^*$ 
at each redshift (De Propris et al. 2003, Haines et al. 2009). General results from the latter approach converge to yield lower values of the blue fractions, rising the suspicion that the BO effect may be partially due to biased photometric selection only. 
In particular De Propris et al. (2003) find that IR-selected blue fractions
are lower than optical-selected ones, and both with no redshift trend when performing the same limiting procedure; they impute the gap to a 
population of starbursting dwarfs or faint spirals
that lie on the border of the variable magnitude limit and hence are missed when selecting in terms of $K^*$.  

We compare these observational data, although diverse in their selection criteria, with our blue fractions derived according to the DS mean colour, 
in Fig. \ref{fblu}. 
Comparing the results obtained in the different colours gives then indications that the BO effect, as originally
introduced, is more evident in optical-optical colours than in optical-NIR: here the blue fraction evolves as more flattened, in agreement also with 
results by De Propris et al. (2003) or Haines et al. (2009). This is expected, since colours bracketing the 4000\AA   break, i.e. with a blue optical colour, are much more sensitive to ongoing star formation, meaning that the physical colour spread will be boosted and the $f_b$ will hence increase the bluer the colour is. 
Our galaxies in {\it B-V} and {\it U-B} fall on a trend fairly compatible with the original definition of BO84, which
is plotted as the curve $f_{bl}=(z/2)-0.015$, extrapolated to high $z$. 
All the observational estimations based on fixed magnitude criteria on datasets at different redshifts, lie instead well above such a line.
We have tested that our SAM curves, when applying such latter selection, are boosted as well up to the same high levels (not shown in plot): in particular they 
would feature the same steep slope at low redshift, followed by a high plateau from $z\gsim$1 (data by Rakos \& Schombert 1995, Fairley et al. 2002, Gerke et al. 2007, Urquhart et al. 2010), that is in contrast with the gentler yet steadier slope extrapolated from BO84 over the same interval. 
On the other side much closer to our models are single cluster
detections at high $z$, such as those by Lerchster et al. (2011) and Fassbender et al. (2014).
A fair explanation to this regard is that two effects play a combined role, enhanced in the visual selection: first a more straightforward one due to incompleteness on pure magnitudes, producing a $f_b$ increase because fewer faint red galaxies are detected in most cases at high $z$. 
Second, an indirect incompleteness effect on colours as well, that causes an increasing $f_b$ towards fainter limiting magnitudes because red-parts are more incomplete:
in fact, since both of any observed bands have a well-defined magnitude limit, the incompleteness increases when moving upwards towards redder colours in the
colour-magnitude-space at a fixed magnitude; this implies that the red galaxy counts are less complete than the blue galaxy counts, resulting again into a systematical increase of the blue fraction.

As to the differences with environments we find that, irrespective of the colour, the blue fraction is higher in groups than in cluster cores (with 
cluster outskirts in between), both in SIM and in SAM: this environmental component of the BO effect was already highlighted in sevaral studies (see e.g. Urquhart et al. 2010).

Finally, we also note that all our SAM curves (and particularly SIM clusters as well) lie lower than another SAM by Menci et al. (2008), who
adopt similar selection criteria as ours in {\it U-B}.

\begin{figure}
\includegraphics[width=8cm,height=6cm]{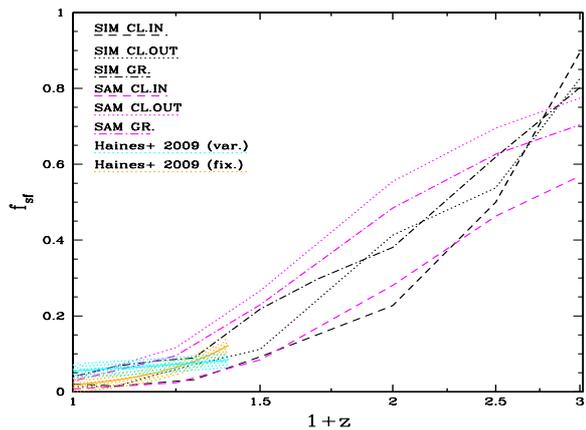}
\caption{Fraction of star-forming galaxies as function of redshift, defined according to a threshold of $sSFR=10^{-10}/yr$.
Data from Haines et al. (2009) at low $z$ yield fractions computed by defining fixed ({\it orange shaded}) or resdshift-variable ({\it cyan shaded}) thresholds
to star-forming activity estimated from $L_{IR}$ and thoroughly converted to SFR assuming a Kennicutt law. 
}
\label{sfr}
\end{figure}

In general, colours provide estimations of star formation histories over longer timescales than usual SFR indicators, keeping also memory of 
past environmental effects; on the other side, the SFR is a more intrinsic property than colours, since it comes as a direct, unprocessed output from models.
Yet when dealing with the modelled parameter, a further caveat is that it refers to the actual SFR in the SAM, while in the simulation it is computed as the average over the last Gyr, so that it traces the galaxy cumulative star formation history more than episodic starbursts.
Also, often in observations more reliable estimates of SFR come from measuring the mid IR re-emission rather than from rest-frame UV/optical data alone
(see Saintonge, Tran \& Holden 2008).
In Fig. \ref{sfr} the fraction of star-forming galaxies is plotted, as computed by imposing a threshold of $sSFR=10^{-10}/yr$ and compared with
{\it K}-band selected data from Haines et al. (2009) at $z\lsim$0.3, that applied either a fixed or redshift-evolving threshold 
on the $L_{IR}$-based SFR (shaded orange and cyan regions, respectively).
Here, our cluster core curves feature a flatter evolution at $z\lsim0.5$ that is consistent with both the observational data sets, followed by a steeper trend thereafter: this is more pronounced in the SIM, where they eventually reach values of star forming galaxies of above 80\% at $z$=2.
For other galaxy classes, the slope begins to steepen at later epochs ($z\sim0.3$) onwards.

All in all, the environmental sequence appears as once again confirmed: inner galaxies are less active than outer and group galaxies. 
However, the curves from different regions in the SIM
tend to converge towards $z>1$, hinting at a reversal of the SF--density relation between cluster cores and outskirts at $z\simeq1.5$, as recently reported by Santos et al. (2015) at exactly that redshift. In the SAM instead the three curves run almost parallel over time, indicating that the environmental
quenching (as measured by the dependence of $f_{bl}$ on the cluster-centric distance) is not changing with redshift, as assessed by Raichoor \& Andreon (2012b).
Finally, the only difference between normal and fossil groups in the SIM (not shown in the figure, where they are unified as average) arises at $z\leq$0.2, 
when FGs abruptly lose all their star-forming galaxies, while NG galaxies keep on their prolongued activity still until present, confirming previous 
findings of R08.
 

\begin{figure}
\includegraphics[width=8cm,height=6cm]{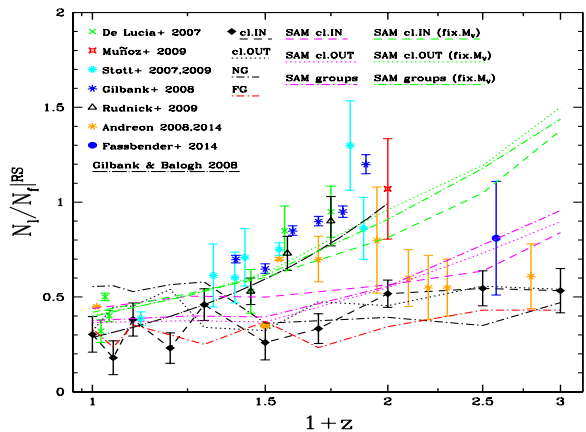}
\caption{Ratio of luminous-to-faint galaxy numbers in the RS sample, defined from bi-colour separation and by fixed mass thresholds
(except the SAM green lines, plotted for reference as to magnitude selection). Errorbars are shown only for cluster cores in SIM.
Dot-long-dashed curve shows the bestfit for $z\leq1$ clusters by Gilbank \& Balogh (2008), using fixed thresholds in magnitude equivalent to $M_V=-20$ and -18.2.
}
\label{lfRS}
\end{figure}

\subsection{The luminous-to-faint ratio}

The LFR (or alternatively the DGR, dwarf-to-giant ratio) is primarily
a function of the LF slope, but is also affected by the value of its characteristic luminosity, hence its estimate is degenerate upon variations 
in both these quantities. Besides, it is also very sensitive to the exact color cut definition, as discussed in the previous section.
The most critical aspect about measuring the LFR is which parts of the LF are considered in the luminous and faint bins. 
In order to make the comparison as much uniform as possible with observational data, we computed 
the luminous-to-faint ratio (LFR) according to different thresholds, either in luminosity or in mass, with the
main purpose of mimicking the completeness limits adopted in most observations.
Since using a fixed threshold in magnitude results into an unrealistic uprise in the LFR at high $z$, 
we choose to separate the two samples either in terms of (fixed) stellar mass, with upper and lower limits given by $M_*=2\cdot 10^{10}$
and $4\cdot 10^{9}M_{\odot}$, respectively; or alternatively in terms of magnitude, that must be consistently evolved with redshift in order to compare galaxies at the same evolutionary stage. To this regard, intervals should be defined with respect to the evolving characteristic magnitude $M^*(z)$ at a given redshift, which is commonly well-approximated with simple stellar population models of formation redshift $z_f=3-4$ (see Andreon et al. 2014). Following e.g. De Lucia et al. (2007), this corresponds to the following interval bins: luminous $\leq M^*+1.2$, $M^*+1.2<$faint$\leq M^*+3$, that in turn translates into the thresholds $M_V(z=0)$=(-20,-18.2) and $M_K(z=0)$=(-23,-21.15), by assuming the local characteristic magnitudes from the previous section.

All the observational points in Fig. \ref{lfRS} are based on cluster surveys in visual wavebands with apertures of few $R_{500}$ at $z\leq1$ 
and mostly display an increasing trend. 
Our results indicate that both SIM and SAM curves are flatter with redshift than most of observations:
in fact, they get the best agreement with Andreon's points (2008, 2014), but also with Fassbender et al. (2014) -these are also among the farthest clusters confirmed in the literature, with various degrees of quiescence measured in their cores. 
Here our RS samples are defined according to the bi-colour separation, but similar results come out from the criterion of DS mean colour.
Regarding environmental effects on the LFR, no such evidence is firmly established from our results, that are thus not able to confirm previous claims
leading to a size or mass segregation, by e.g Ba\~nados et al. (2010) and de Filippis et al. (2011).
As further test, the SAM curves obtained applying a fixed magnitude criterion for selection are plotted alongside in the same figure (green lines):
as expected, they instead trail on the steep trend fitted by the SAM in Gilbank \& Balogh (2008), valid for clusters up to $z\simeq$1. 
This confirms that the LFR evolution is mostly sensitive to the selection criteria chosen.

\begin{figure}
\includegraphics[width=8cm,height=5cm]{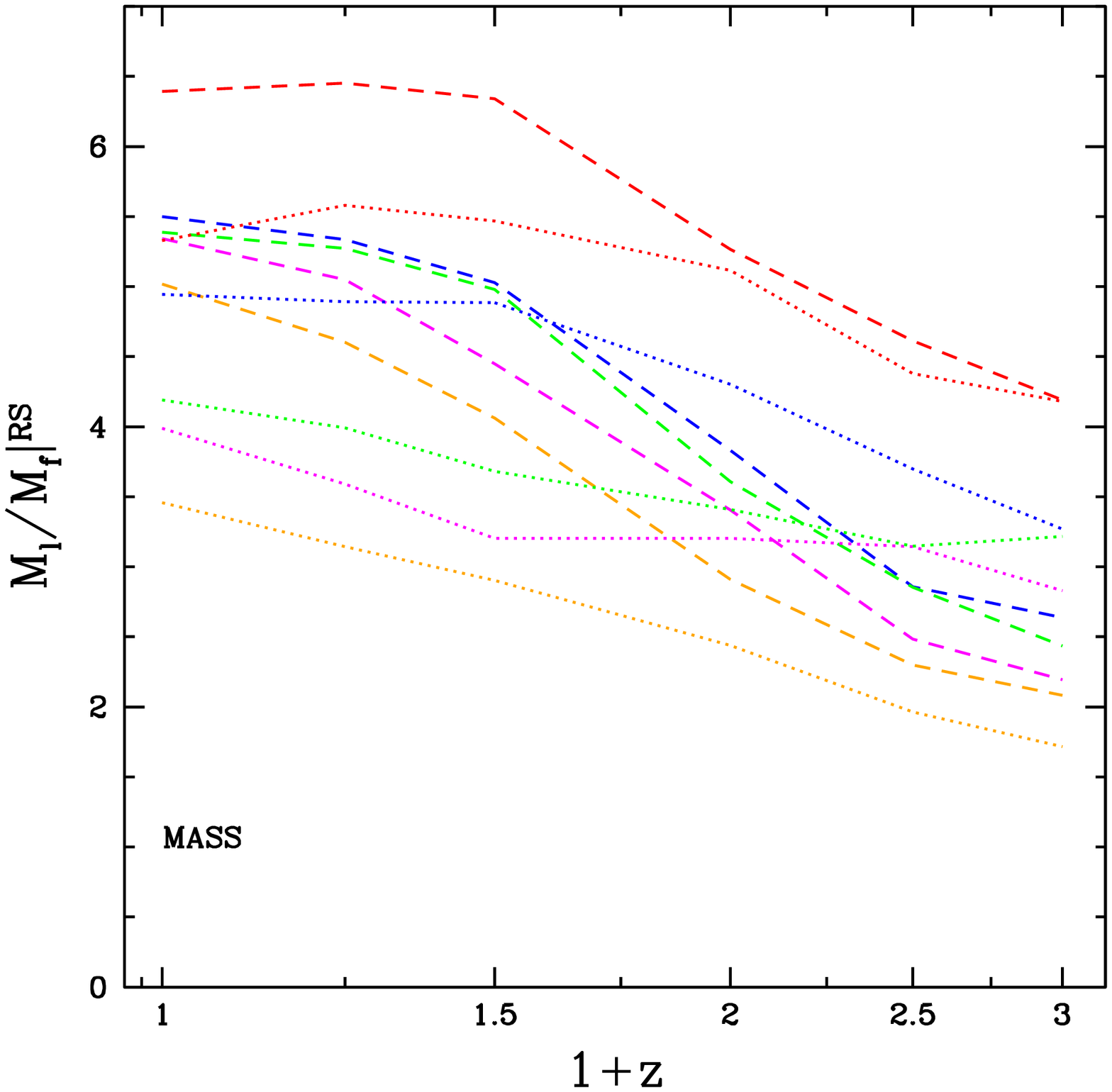}
\includegraphics[width=8cm,height=5cm]{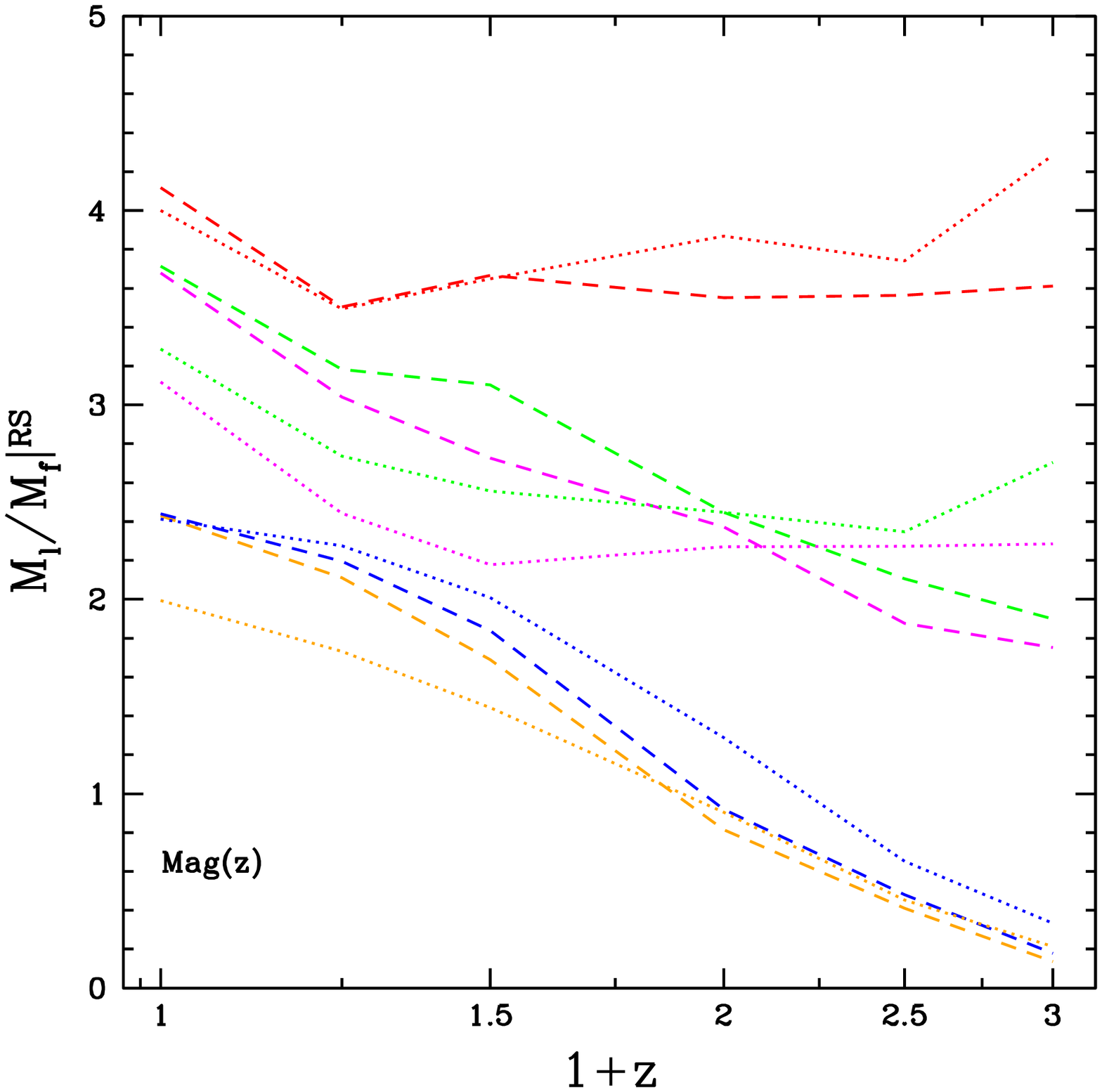}
\includegraphics[width=8cm,height=5cm]{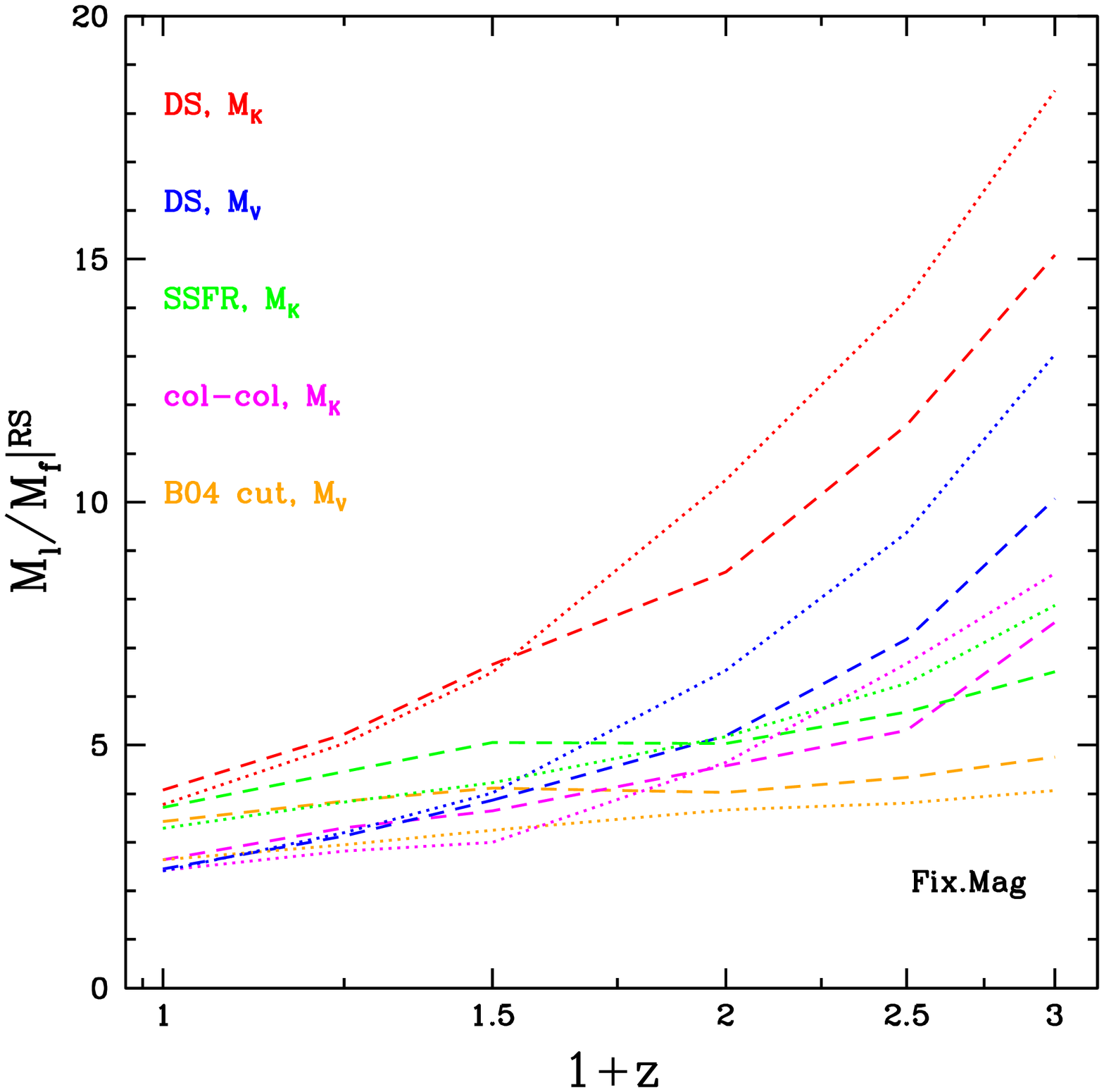}
\caption{Ratio of stellar masses contained in luminous-to-faint RS galaxies in the SAM for cluster cores ({\it dashed}) and averaged cluster outskirts plus groups  ({\it dotted}), according to the different selection criteria listed. 
{\it Up}: Fixed mass limits at $M_*=2\cdot 10^{10}$ and $4\cdot 10^{9}M_{\odot}$; {\it middle}: redshift-variable magnitude limits (see text); {\it bottom}: fixed magnitude limits at $M_V=$-20 and -18.2, or $M_K=$-23 and -21.}
\label{mrRS}
\end{figure}

The sensitivity of the LFR on the sample selection function is even more evident if the ratio of stellar masses is considered instead of the 
galaxy numbers, although this cannot be considered as a proper observable; in Fig. \ref{mrRS} SIM are not shown because of their poorer statistics, 
resulting in larger errors expecially at high $z$, since the mass ratio is strongly affected by even negligible shifts in galaxy numbers.
In the mass-selected sample (upper panel) an increasing trend with time is followed by all SAM curves: red galaxies with stellar masses
greater than $2\times 10^{10}$ grow their stellar mass by approximately twice especially in cluster cores and for almost all the colour selection
criteria, although with different normalizations. 
On the opposite, when the lower limit is set in terms of the same fixed magnitude used in De Lucia et al. (2007) or Gilbank \& Balogh (2008), both aiming at reproducing
the original procedure by BO84 but with fixed limits, the picture gets upside down and the stellar mass contained in brighter galaxies decreases
in time with respect to that in fainter ones (lower panel).
The latter behaviour is more consistent with a rapid buildup at the low-mass end of the quiescent stellar mass function, as measured for example by Tomczak et al. (2014) in the field, who found that the total stellar mass density of passive galaxies (complete down to $10^9M_\odot$) has increased since $z$=2.5 by a factor of 12, whereas that of star-forming galaxies only by about twice.
Finally, an intermediate behaviour between the two previous ones is displayed in the mid panel, where the limits are given in terms of evolving magnitude as
aforementioned: in this case the overall trend of the mass ratio is roughly constant over time, especially when measured in {\it K} band, while in {\it V} it approximately doubles from $z$=2 to present (yellow and blue lines). We tend to favour the latter scenario, as well as the one descending from the mass selection, because more physically motivated since its premises.

\section{Discussion and Conclusions}

In this work we have compared theoretical data from two different sources, a cosmological-hydrodynamical simulation and a semi-analytical model, to observational data. 
We analyzed proxies of galaxy populations in clusters and groups such as blue fractions and LFR on the RS, extending the analysis over a wide range of optical and NIR wavelength. Both these observables correspond to two major controversial outputs from any galaxy formation model, namely how many red and how many dwarf galaxies are formed with respect to the observed local and high-redshift stellar mass functions.
We applied different criteria for defining blue and red galaxies, all redshift-dependent in order to disentangle the natural colour reddening from the intrinsic evolutionary effects we want to study.
Likewise, we selected our RS samples by defining limits in both $M_V$ and $M_K$ that are evolving with redshift, to take into account the expected luminosity evolution of galaxies, as parametrized by $L^*$ at each $z$. The combination of these selection criteria ensures to account for the naturally bluer rest-frame colours of high-redshift galaxies, given their mean younger age and the higher mean SFR of the overall Universe at that epoch.
Our main findings are the following:

\begin{itemize}
\item{The blue fraction's evolution is more marked when selecting in optical colours than in {\it K-}band. 
Visual blue fraction (Fig. 3) as well as star-forming fraction (Fig. 4) evolve more steeply earlier than $z\simeq0.5$, especially in SIM. 
The much slower evolution at $z\lsim0.5$ is in good agreement with data by De Propris et al. (2003) or Haines et al. (2009). 
At high redshift our results are consistent with spectroscopically confirmed proto-clusters with deep completeness in galaxy mass such as
Fassbender et al. (2014) and Andreon et al. (2014).}
\item{Trends in blue fractions are definitely flatter when limiting in stellar mass or evolving magnitude, rather than fixed {\it V} magnitude (Fig. 3).}
\item{LFR trends are compatible with the most recent data at high redshift (e.g. Andreon et al. 2014, Fassbender et al. 2014) pointing at very mild evolution, either when selecting by mass or by (evolving) magnitude (Fig. 5).}
\item{Both results of blue/red fractions and LFR are inconsistent with observations that applied fixed thresholds in limiting magnitude or colour cut 
upon redshift-spanning surveys, but do instead fairly reproduce their steeper evolution when adopting the same criteria.}
\item{Testing the different criteria adopted for defining RS galaxies, our conclusion is that choosing a physically motivated redshift-dependent colour
threshold is of capital importance. However among the three sources of uncertainty that we analyzed, the observable results are affected first by the fixed/variable magnitude choice on the lower limit, secondly by the magnitude band itself (visual or {\it K}) and thirdly by the colour criterion among those ascertained (see Fig. 6).}
\item{Normal groups and cluster outskirts present higher blue and star-forming fractions and LFR than cluster cores at any epoch, consistently with a picture where groups are not very efficient at quenching infalling galaxies (see Gerke et al. 2007). When looking in particular at FGs, our results support previous findings by Romeo et al. (2008), that the number of
blue galaxies begins to deviate with respect to that of normal groups only at $z\lsim0.5$ (Fig. 3), due to a much larger fraction of red dwarfs in FGs (Fig. 5).
This confirms again that star formation history in FGs tends to behave close to cluster cores during the last 4 $Gyr$ of cosmic time, after having followed
the same pattern as normal groups during the previous epoch.}
\item{A source of discrepancy between SIM and SAM is given at early epoch: cluster cores in SIM display evidence of higher SFR in the past earlier than 
$z\sim$1.5, in agreement with clues from recent detections (e.g. Strazzullo et al. 2013, Santos et al. 2015); on the contrary the redshift dependence
of $f_{bl}$ in SAM keeps the same slope among the three galaxy classes, varying only its normalization, and pointing at a redshift-independent environmental quenching during the time interval considered. This difference could root into the epoch of stellar mass assembly, that is generally anticipated in the SAM with respect to SIM (see below).}
\item{By adopting a mass-selected sample we also find that the more massive RS galaxies grow their mass at a higher rate than less massive ones (Fig. 6): 
this reflects that progressively more massive former blue galaxies keep joining the RS once quenched (cf. Faber et al. 2007, Cen 2014).
We remark that since these are selected as already quiescent galaxies, they are likely to grow in stellar mass through (minor and dry) merger events after
entering the RS, thus contributing to steepen its slope (see De Lucia et al. 2006, Romeo et al. 2008, Guo \& White 2008, Jim\'enez et al. 2011). }

\end{itemize}

We note that in all the plots the  differences between SIM and SAM are always much smaller than the typical scatter of the observed data.
In general SAMs produce an excess of stellar mass in early low-mass galaxies, due to an over-production of stars in low-mass DM haloes,
and along with this they result into a too high fraction of red satellites in clusters (see Fontanot et al. 2009, Guo et al. 2011). 
Both these manifestations of the same phenomenon are due to galaxies that form very early and evolve almost passively at later times, because following 
too closely the hierarchical growth of their DM host haloes: to this respect a way to overcome such drawback could be decoupling the halo accretion rate 
from the galaxy SFR (see Weinmann et al. 2012).
Various {\it ad hoc} models have been proposed to quench early star formation in dwarf galaxies:
for example by halting further gas accretion below a certain halo mass threshold (Bouché et al. 2010), or by introducing a two-phase gas split
into star-forming and not (Cousin et al. 2015), in order to reduce the star formation efficiency in low-mass haloes since early epoch.
However in the range of mass below $10^{10} \, M_{\odot}$, physical processes such as feedback by SNe and stellar winds play a crucial role on galaxy evolution, 
that are still likely implemented in a too simplistic way into the models (see Kannan et al. 2014).

In particular Weinmann et al. (2011) found a red fraction of modelled galaxies too high when compared to Virgo cluster or more generally to
less dynamically relaxed clusters such as higher-redshift systems,
ascribing it to either an over-efficient quenching of star formation in satellite galaxies (as found by Guo et al. 2011), or to the lack 
of an efficient stellar stripping/disruption of such galaxies. 
Nevertheless, the latter channel has been found not to represent a complete solution: in fact, recently Contini et al. (2014), who updated the SAM of 
De Lucia \& Blaizot (2007) with specific prescriptions for stellar stripping, addressed the problem of the over-prediction of low-mass 
galaxies and demonstrated that it is only partially alleviated thereby, since tidal forces preferentially act on intermediate to high-mass galaxies. 
Therefore all present SAMs are facing the same difficulties, given that they normally assume that satellite galaxies get quenched in all haloes regardless 
of mass or redshift.

For what concerns the differences between SIM and SAM, these mainly arise at $z\ge1.5$, where simulations yield higher SFR (Fig. 4) and lower LFR (Fig. 5).
In addition SAM tends to yield redder colours at any epoch, even when considering only quiescent galaxies and especially in the IR (Fig. 1).
In particular the bulk of star formation in SAM core galaxies occurs fairly earlier than $z$=2, whilst these same objects in SIM are still actively forming
stars at that epoch. This discrepancy at early epoch can lso originate from the more virialized dynamical state of the averaged sample of SAM haloes.

Our models differ, besides resolution effect, chemical enrichment (which is instantaneous in the SAM), diffuse light (which is present as inter-galactic star
particles in the SIM, see Sommer-Larsen, Romeo \& Portinari 2005) and possibly different dwarf modelling, also for the fact that the simulation does not implement AGN feedback, while the semi-analytical model does: here star formation in the central galaxies is shut off when radio-mode AGN activity is turned on and prevents halo gas from cooling on to the galaxy itself.
Therefore SFR in SAM core galaxies can be lower than in SIM, as some massive satellite galaxies are already quenched by AGN feedback before accretion.
There is nowadays a significant evidence of AGN feedback just at
the epoch of interest to the present work, driven by IFU-observed molecular outflows (see e.g. Brusa et al. 2015, Cresci et al. 2015, Perna et al. 2015). However, from our comparison we see that models including or excluding AGN feedback give comparable results as far as the colour distributions and fractions 
and the stellar mass function are concerned, provided that only galaxy samples excluding the BCGs are considered. 
On the opposite mass side, our joint results from SIM and SAM converge in not relating any proper dearth of red dwarf galaxies above that epoch, although with
the double caveat that detection of red satellites is observationally challenging at $z\gsim$1, and that especially SAMs usually meet severe difficulties
in reproducing the correct stellar mass function at its faint end.

The emerging picture that we gather is that the main physical mechanisms acting in either of our models, namely cold gas removal by ram pressure stripping on one side, whose efficiency increases in denser environments, and AGN feedback on the other, which affects the more massive galaxies by virtue of their halo mass,
can combine together, and with the aid of a long lasting sequence of minor/dry merging activity, to shape the quenching history of galaxies as a function of both their environment and stellar mass. 

\section*{Acknowledgments}

We thank the referee and also S. Andreon for their useful comments.
This work is supported by the 973 program (n. 2015CB857000, 2013CB834900), the Foundation for Distinguished Young Scholars of Jiangsu Province (n. BK20140050), the ``Strategic Priority Research Program {\it The Emergence of Cosmological Structure}'' of the CAS (n. XDB09010000) and by China Postdoctoral Science Foundation through grant n. 2015M570488 (AR). 
AR and XK are also supported by the NSFC (n. 11333008). EC acknowledges the support from the CAS Presidents International Fellowship Initiative (n. 2015PM054).
RF acknowledges funding from the European Union Seventh Framework Programme (FP7/2007-2013) under grant agreement n. 267251 `Astronomy Fellowships in Italy' (AstroFIt).

\end{document}